\documentclass[twocolumn]{revtex4}
\usepackage{amssymb,amsmath,amsfonts,bm}
\usepackage{graphicx}
\usepackage{color}

\newlength{\plotwidth}
\begin{document}

\date{\today}

\title{Cascades in nonlocal turbulence}

\author{Gregory Falkovich$^{1,2}$ and Natalia Vladimirova$^3$}
\affiliation{
$^1$Weizmann Institute of Science, Rehovot 76100 Israel\\
$^2$Institute for Information Transmission Problems, Moscow, 127994 Russia\\
$^3$University of New Mexico, Department of Mathematics and Statistics, Albuquerque NM 87131
}

\begin{abstract}
We consider developed turbulence in the 2D Gross-Pitaevsky model, which
describes wide classes of phenomena from atomic and optical physics to
condensed matter, fluids and plasma.  The well-known difficulty of the
problem is that the hypothetical local spectra of both inverse and
direct cascades in the weak-turbulence approximation carry fluxes which
are either zero or have the wrong sign; such spectra cannot be
realized. We analytically derive the exact flux constancy laws
(analogs of Kolmogorov's 4/5-law for incompressible fluid turbulence),
expressed via the fourth-order moment and valid for any nonlinearity.
We confirm the flux laws in direct numerical simulations. We show
that a constant flux is realized by non-local wave interaction in both
the direct and inverse cascades. Wave spectra (second-order moments) are
close to slightly (logarithmically) distorted thermal equilibrium in
both cascades.

\end{abstract}


\maketitle


Turbulence is a state where pumping and dissipation happen at very
different scales, so that the main issue is the nature of the transfer
of a conserved quantity (say, energy) from pumping to damping due to
nonlinear interaction.  Despite the fundamental importance of this
process, there is a certain terminological (and even conceptual)
confusion surrounding the notion of locality of turbulence cascades.
It may seem clear intuitively: either the energy is transferred
locally in $k$-space by a cascade-like process or jumps directly from
the motions of pumping scales to those of damping scales.

The problems start when one tries to formally sort out how locality or
non-locality is manifested in the correlation functions.  The simplest
case is that of a complete scale invariance when the statistics at the
scales between the pumping scale $l_p$ and the damping scale $l_d$ are
independent of $l_p$ and $l_d$; it is then natural to call it a local
cascade.  Such cases do exist at least for some inverse cascades;
however, scale invariance is spontaneously broken in direct cascades,
see e.g.~\cite{Falkovich2009}.  Indeed, the direct energy cascade in
3D has $l_p$ explicitly entering all the velocity structure functions
except the third, which determines the energy flux through the scale
$r$.  The same is true for the direct vorticity cascade, where all
velocity and vorticity correlation functions contain $l_p$, again,
except the triple moment expressing the flux of squared
vorticity~\cite{kraichnan1967condensate,kraichnan1967inertial,FL94}.
Are we to call such turbulence non-local, because the pumping scale
explicitly determines the moments (including the second one, i.e. the
energy spectrum), or to classify it as a local cascade, because the
flux through the scales is constant?

One finds even more confusion in weak turbulence theory, which aspires
to provide a close description solely in terms of the second moment --
if it contains $l_p$ and/or $l_d$, turbulence is called
non-local~\cite{ZakharovLvovFalkovich1992}. We wish to state here that
even in such cases there must exist a higher-order correlation
function which is universal, i.e., dependent neither on $l_p$, $l_d$ nor
on mechanisms of pumping and dissipation.

It is thus important to clearly distinguish non-locality of turbulence
(when some correlation functions in the inertial interval depend on
the pumping and/or dissipation scales) and locality of the flux, which
corresponds to a single correlation function independent of $l_p$ and
$l_d$.  Locality of the flux is, in a sense, a trivial consequence of a
conservation law, and it may co-exist with the non-locality of turbulence.

Here we show that such co-existence takes place for 2D turbulence in
the framework of the Gross-Pitaevsky model. We derive universal
(i.e., independent of $l_p$, $l_d$) flux relations for both direct and
inverse cascades. These universal flux relations are expressed via the
fourth moments. And yet we show that both cascades are not
scale-invariant, so that the second moments (and spectral densities)
explicitly depend on the pumping and dissipation scales.


The Gross-Pitaevsky model is one of the most universal models in
physics. It describes wave propagation in wide classes of phenomena in
fluids, solids and plasma.  Applications include light propagating in
media with the Kerr nonlinearity~\cite{SulemSulem1999}, and
non-equilibrium states of cold atoms in Bose-Einstein
condensates~\cite{PitaevskiiStringariBook2003}.

The range of applications is broad because the physical model is built
on a single assumption: the narrow distribution in the space of momenta of
wave vectors.  The complex wave envelope, $\psi$, then evolves according
to the Gross-Pitaevsky, or nonlinear Schrodinger equation,
\begin{equation}
  \psi_t=i\nabla^2 \psi + is |\psi|^2\psi.
  \label{NLSE}
\end{equation}
On the right-hand side, the first term describes a linear propagation,
and the second term represents nonlinear interaction (of waves or
particles).  The parameter $s$ in front of the nonlinear term
distinguishes the focusing/attractive ($s=+1$) and
defocusing/repulsive ($s=-1$) cases.

Equation \eqref{NLSE} has two integrals of motion (conserved
quantities): the wave action, ${\cal N}=\int|\psi|^2d{\mathbf r}$, and the
Hamiltonian, ${\cal H} = \int |\nabla \psi|^2 - \frac{1}{4} s |\psi|^4
d{\mathbf r}$.  Assuming a weak nonlinearity, it can be shown that the cascade
of wave action in spectral space is
inverse~\cite{ZakharovLvovFalkovich1992}, i.e.,  directed toward small
wave numbers, while the cascade of the energy is direct.
Independent of the sign of the nonlinearity, the inverse cascade results
in the appearance of large, spatially-coherent
structures~\cite{dnpz,DyachenkoFalkovich1996}, and, in the case
of the defocusing nonlinearity, in the accumulation of condensate --- the
mode that is spatially-coherent across the whole system.  In this study,
however, we avoid condensate creation by adding dissipation at low
$k$. Therefore, even though we work with $s=-1$, some of our results
could be relevant to the focusing case as well.

We numerically solve Eq.~\eqref{NLSE} with the defocusing
nonlinearity, using a standard split-step method~\cite{dnpz} modified
to be 4th-order accurate in time.  Forcing and damping, applied in
spectral space, are represented in the r.h.s. of the equation,
\begin{equation}
   i\psi_t + \nabla^2\psi - |\psi|^2\psi = i\hat{f_k} \psi + i\hat{g_k}.
  \label{NLSEclean}
\end{equation}
We use the same code as in~\cite{vlader2012}, with the numerical method
described in detail in~\cite{arxiv2011}.  Our computational domain is square,
$L\times L$, with periodic boundary conditions, so that the lowest
wave number is determined by the domain size, $k_{\min} = 2\pi/L$.
All simulation presented here are done at the same resolution, $\Delta
x = \Delta y = 2\pi/1024$, and time step $\Delta t \le 10^{-5}$.  We
use domains up to $L=32\pi$, with grids up to $16386^2$ points.

\begin{figure}
\begin{center}
\includegraphics[width=80mm]{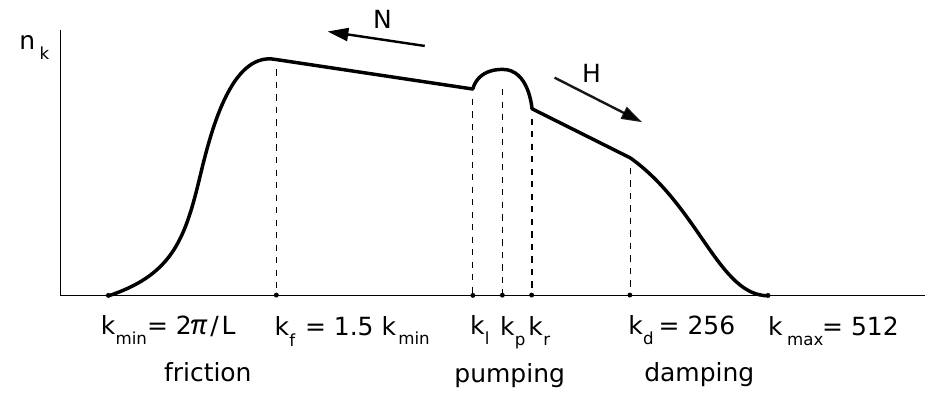}
\end{center}
\caption{Schematic representation of forcing and damping in spectral space.}
\label{fig:setup}
\end{figure}

To deposit wave action into the system we use the additive forcing,
$g_k = |g_k| e^{i\phi_k}$, with random phases $\phi_k$ and amplitudes
$|g_k| \propto \sqrt{(k^2 - k_l^2)(k^2_r - k^2)}$ which are non-zero
only in a ring of wavenumbers $k \in [k_l,k_r]$.  The forcing is
normalized to deposit a specified amount of wave action, $\dot{N}
\equiv \alpha$, where $N= \langle|\psi|^2\rangle$ and angular brackets
denote spatial averaging.  The multiplicative forcing, $f_k = - \beta
(k/k_d)^4(k/k_d - 1)^2$, provides small scale damping at $k> k_d$.  We
use $k_d \approx 3 k_r$ to include a contribution of cubic nonlinearity
to the direct cascade.  In addition to high-$\mathbf{k}$ damping, we have
an option to include low-$k$ friction, $\, f_k =
-(1,1,\frac{1}{\sqrt{2}})\, \gamma\,$ for $\,k = (0,1,\sqrt{2})
k_{\min}$. Unless otherwize specified, we use $\gamma =
6.2\alpha^{\frac{2}{3}}$.  To distinguish dissipation at low and high
wavenumbers we will refer to them respectively as ``friction'' and
``damping'' to stress that they play different roles.  The purpose of
friction is to absorb the wave action and prevent the accumulation of
condensate, while the purpose of damping is to absorb the energy and
limit the spectrum to a finite number of modes.

One advantage of running at the same resolution is that we can use the
same damping parameters, $k_d=256$ and $\beta=400$, in all
simulations.  Our studies of the inverse cascade are done with pumping
rings with $k_l = 68$ and $k_r = 84$, while our simulations of the direct
cascade are done with $k_l = 6 k_{\min}$ and $k_r = 9 k_{\min}$.  The
strength of forcing and friction are controlled by $\alpha$ and $\gamma$
respectively, which are treated as simulation parameters.

Notice that $\alpha$ is an input rate of wave action into the system,
which is close to but not exactly equal to the flux into the inverse
cascade, $\tilde \alpha$.  In our simulations of the inverse cascade, at
most 10\% of wave action is lost to damping, that is $\tilde \alpha
\gtrsim 0.9 \alpha$.  The lost fraction is measured by computing the
wave action consumed by friction in a steady state, and also from
simulations without friction, where the slope of $N(t)$ initially
matches the rate of wave deposition $\alpha$ and switches to
$\tilde{\alpha}$ later.



We start with the demonstration that the flux laws expressed in terms of
fourth moments are exact and local, that is, independent of $k_p$ and
$L$.
%
Let us consider steady-state turbulence and look at the following quantity,
$\langle |\psi_1 - \psi_2|^2\rangle$.  Here, $\psi_1$ and $\psi_2$
refer to the values at two points separated by the distance $r$ taken
in the interval of the inverse cascade, $L\gg r\gg k_p^{-1}$.
This second moment can be related to the spectral density:
\[
  \langle |\psi_1 - \psi_2|^2\rangle = \int |\psi_k| ^2(1-\cos k r ) dk
        \simeq \int_{1/r}^\infty  | \psi_k |^2 dk.
\]
Next, we use Eq.~\eqref{NLSEclean} to take the time derivative of
\[
   \langle |\psi_1 - \psi_2|^2\rangle =
       2N - \langle \psi_1\psi_2^* + \psi_1^*\psi_2\rangle
\]
and obtain,
\begin{equation}
       Q(r) \equiv 2 \, {\rm Im} \langle \psi_1^*|\psi_2|^2\psi_2 \rangle =
       -\tilde\alpha.
  \label{eq:qflux}
\end{equation}
This exact flux relation is valid for any dimensionality.  Note that
the right hand side is an outcome of all mechanisms of pumping and
dissipation acting at the scales smaller than $r$, i.e., the effective
flux of the wave action into the inverse cascade. This result is
non-trivial because it shows that the fourth correlation function
$Q(r)$ is the divergence of the wave action flux, which is equal to
the input rate in the steady state, and thus does not depend on the
distance $r$.

\begin{figure}
\begin{center}
\includegraphics[width=\plotwidth]{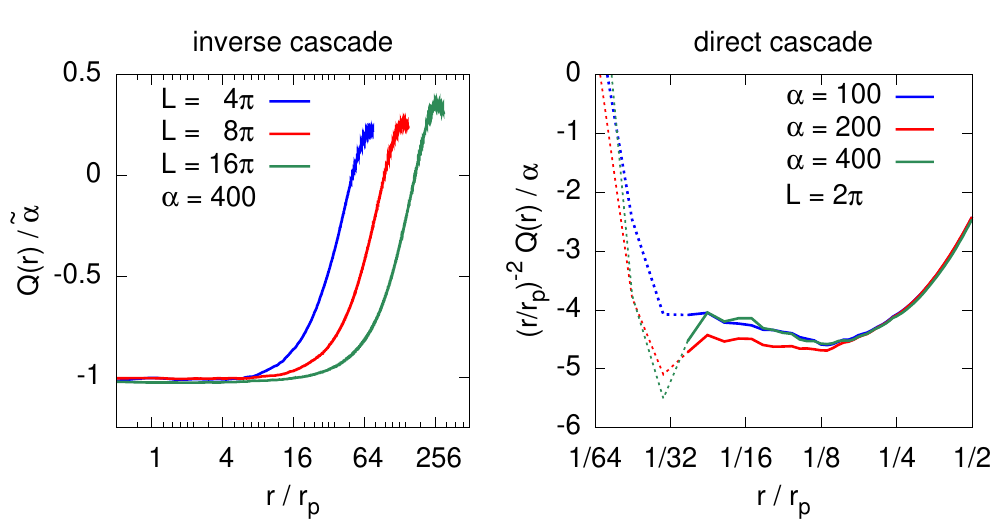}
\end{center}
\caption{
  The ratio of the correlation function $Q(r)$ to the input rate of
  wave action in simulations of inverse and direct cascades (left and
  right panels respectively).  The data are collected over time and
  averaged in angular direction. The filter removing harmonics with $k
  > k_l$ is applied before averaging in the inverse cascade runs.
  Dashes lines are data in the damping interval of scales. (Color online.)
}
\label{fig:qflux}
\end{figure}

Simulations with different input rates show that $Q(r)$ scales as
$\alpha$ in the range from the damping scale to the size of the
domain, for both the direct and inverse cascades.  Simulations of the
inverse cascade confirm the appearance of the plateau, $Q(r) =
-\tilde{\alpha}$ for $r \gtrsim r_p$, as shown in
Fig.~\ref{fig:qflux}, left panel.  Here, we have filtered out
small-scale oscillations $r < r_p$, resulting from spectrally narrow
pumping.

For the direct energy cascade, one generally cannot obtain a
scale-invariant flux law since the energy contains two terms:
the quadratic kinetic term $|\nabla\psi|^2$ and the quartic potential term
$|\psi|^4$. At least for not very strong nonlinearities, when
potential energy is not very large, we expect the kinetic energy flux,
$P = \nabla^2 Q$, to be scale-independent, or equivalently $Q(r) \sim
r^2$.  The right panel of Fig.~\ref{fig:qflux} shows that indeed $Q(r)/\alpha
\approx -4.5 (r/r_p)^2$, resulting in $P
\approx -18 \, \alpha r_p^{-2}$,  for $r \lesssim 0.2\,r_p$.

Note that our simulations of systems evolving freely,
without friction, show that there is no time interval when the flux
$Q$ is constant in the inverse cascade, while the region $Q\propto
r^2$ is formed in the direct cascade region. Thus, the flux law
(\ref{eq:qflux}) is specific to steady spectra. This by itself is
already a signature of non-locality --- indeed, in incompressible
turbulence, the 4/5 law of the direct cascade takes place even for
decaying turbulence, while the 3/2 law of the inverse cascade takes
place at a given scale when larger scales continue to evolve, see
e.g.,~\cite{frisch,falgaw2014}.



\begin{figure}
\begin{center}
\includegraphics[width=\plotwidth]{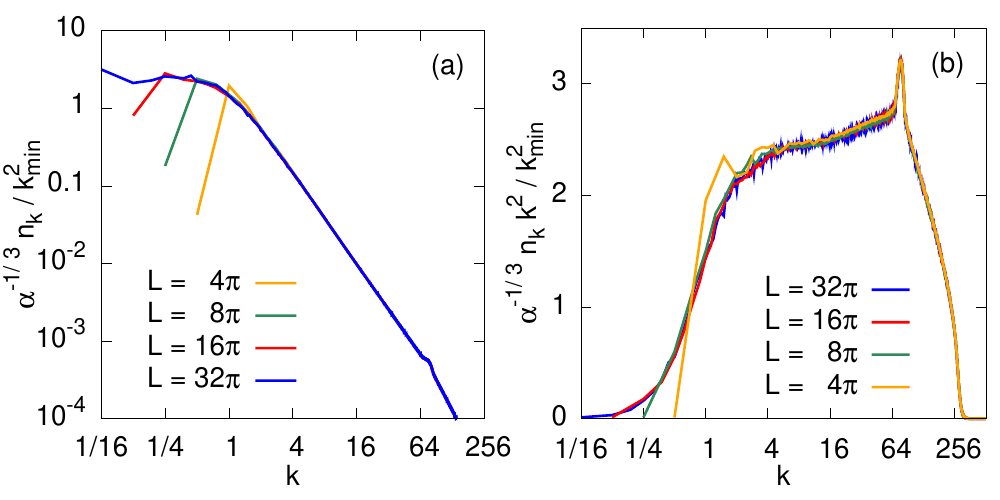}
\caption{
  Non-compensated (a) and compensated (b) spectra in
  simulations with different domain sizes with $\alpha = 400$.
  (Color online.)
}
\label{fig:kmin}
\end{center}
\end{figure}

Next, we look at the spectra, starting with the inverse cascade.  As
was observed in
~\cite{dnpz,DyachenkoFalkovich1996,nazono2006,sun2012}, such spectra
are close to thermal equilibrium, $n_k = T/(k^2 + k^2_\mu)$, where the
temperature $T$ and chemical potential $\mu \equiv k^2_\mu$ are system
dependent parameters.  The transition to the equipartition region,
$k_\mu$, is independent of the size of the domain, as shown in
Fig.~\ref{fig:kmin}.  When the range of $k$ is limited, there is no
equipartition region at all.

\begin{figure}
\begin{center}
\includegraphics[width=\plotwidth]{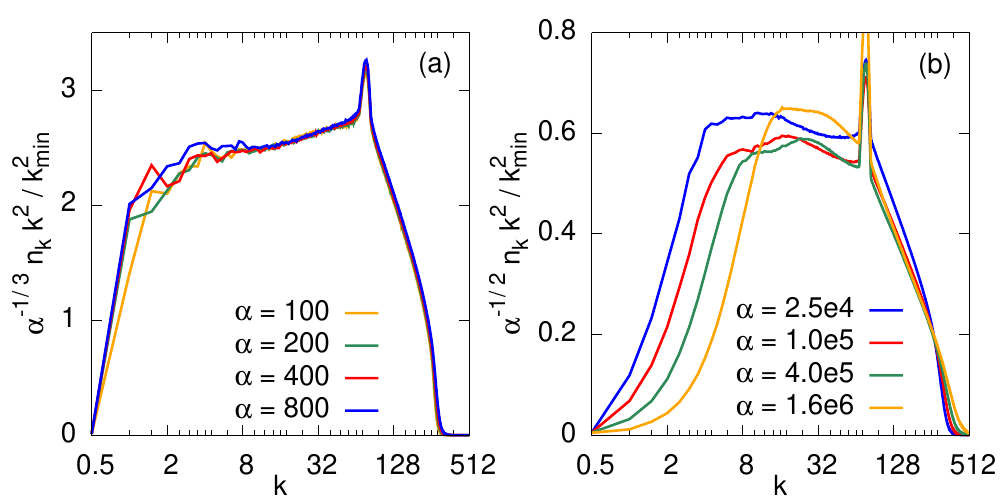}
\caption{
  Effect of pumping rate on stabilized spectra in simulations with
  $L=4\pi$.  At lower rates $n_k \sim \alpha^{1/3}$ (a),
  while at higher rates $n_k \sim \alpha^{1/2}$ and spectra develop
  equipartition region (b).  Friction was selected to
  minimize the pile-up: $\gamma = 6.2 \alpha^\frac{2}{3}$ for $\alpha
  \le 800$ and $\gamma = (250 \alpha)^\frac{1}{2}$ for $\alpha \ge 2.5
  \times 10^4$. (Color online.)
}
\label{fig:alpha}
\end{center}
\end{figure}

We characterize the degree of nonlinearity by the ratio of the mean
potential energy of interaction to the kinetic energy:
$H_p/H_k=\langle |\psi|^4\rangle/\langle|\nabla\psi|^2\rangle$. In
simulations with smaller nonlinearities ($100 \le \alpha \le 800$,
$0.07 < H_p/H_k < 0.15$) the spectra have the weak-turbulence scaling,
$n_k \propto \alpha^{1/3}$ \cite{ZakharovLvovFalkovich1992},
confirming dominance of resonant four-way interactions (see
Fig.~\ref{fig:alpha}a).  The higher-$k$ parts of compensated spectra
(to the immediate left of pumping) have a well-defined slope, which
can be described by a logarithmic correction.  This part of the
spectrum is relatively insensitive to friction.  There is a pile-up at
low $k$; the amplitude and location of the pile-up depend on the
friction.  Adjusting $\gamma$, we can minimize the pile-up but cannot
eliminate it completely.


The pile-up is more pronounced at higher nonlinearities ($2.5 \times
10^4 \le \alpha \le 1.6 \times 10^6$, $0.4 < H_p/H_k < 1.5$)
eliminating the part of the spectrum that logarithmically decreases
towards lower $k$ (Fig.~\ref{fig:alpha}b).  At high nonlinearities we
start to observe a region with action equipartition.  The gap at low
$k$ in Fig.~\ref{fig:alpha}b is similar to the gap in
Fig.~\ref{fig:kmin}b; the higher $\alpha$, the wider the gap.  Unlike
the amplitude of the pile-up, which is very sensitive to friction, the
width of the gap is relatively robust; the friction only slightly
affects the width of the equipartition region.

At high input rates the overall level of spectra changes from $n_k
\propto \alpha^{1/3}$ scaling to $n_k \propto \alpha^{1/2}$, suggesting
transition to three-way interactions, which are non-resonant without a
condensate.  It could well be that in our case the role of condensate
(or ``pre-condensate'') is played by the collection of low-$k$
modes. To analyze the degree of coherence between these modes and
their role in interactions with the rest of the spectrum remains a
subject for future work.

A pile-up and low-$k$ equipartition, similar to those shown in
Figs.~\ref{fig:kmin} and~\ref{fig:alpha}, was observed in simulations
of water wave turbulence and attributed to the bottleneck \cite{bot}
due to lack of low-frequency modes~\footnote{A. Korotkevich, private
  communications}.  In our case, we can rule out this explanation,
since our simulations performed in increasing domains show the
appearance of the equipartition region even for moderate $\alpha$.
Note briefly that the form of the spectrum deviation suggested
in~\cite{dnpz} is not supported by our data.

\begin{figure}
\begin{center}
\includegraphics[width=\plotwidth]{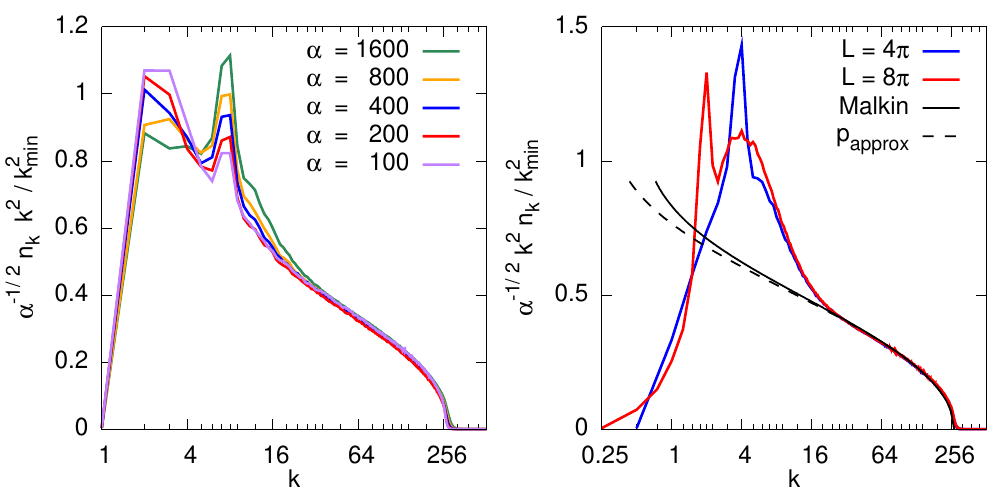}
\caption{
  Left: Spectra of direct cascade for $L=2\pi$ and different $\alpha$.
  Right: Spectra with $\alpha=400$ compared with Malkin model, fit
  with $C=88$. (Color online.)
}
\label{fig:spectra_malkin}
\end{center}
\end{figure}

Next, we describe the steady-state spectra of the direct cascade,
shown in Fig.~\ref{fig:spectra_malkin}.  Far from the pumping, the
spectra scale as $n_k \propto \alpha^{1/2}$, similarly to inverse
cascades with high nonlinearities.  As we extend our computational
domain, we observe a non-universal region at low $k$ and a universal
region at high $k$ (in our case, $k>16$).  We compare the shape of the
spectra in the universal region to the theory by
Malkin~\cite{malkin1996}. The theory assumes non-local
weakly-nonlinear interaction and describes the shape of direct cascade
in terms of $N_k/N$, which is the fraction of wave action contained
within a sphere of radius $k$.  The spectrum, $n_k$, is described
implicitly through the following equations,
\begin{eqnarray}
 &&  \frac{n_k k^2}{k^2_{\min}} = \frac{C}{2\pi} \left[ \ln \frac{N}{N_k}\right]^{\frac{1}{3}},
 \label{eq:Malkin1}\\
 &&   \frac{C}{N} \,\ln\frac{k_M}{k} =   p\!\left( \frac{N_k}{N} \right),
 \label{eq:Malkin2}
\end{eqnarray}
where $C$ is a constant related to the energy flux, and $k_M$ is the
cut-off mode.  The function $p(m)$ is an integral which can be
expressed in terms of the lower incomplete gamma function, ${\cal
  P}(a, x)$, e.g., Eq.~(6.5.1) in~\cite{abramowitz1972},
\begin{equation}
 p(m) = \int_m^1 \textstyle \ln^{-\frac{1}{3}} \! \frac{1}{y} \,dy  =
\Gamma \! \left(\frac{2}{3}\right) {\cal P}\!\left(\frac{2}{3}, \, \ln \frac{1}{m} \right).
\label{eq:Malkin3}
\end{equation}

To compare our data and the theory predictions, we extract $N_k/N$
from numerical simulations and verify Eq.(\ref{eq:Malkin1}) and
Eq.(\ref{eq:Malkin2}) individually.  The comparison shows a good
agreement in the range $\ln(N/N_k) \in [10^{-3}, 0.3]$, or for $N_k/N
\in [0.74, 0.999]$.  We stress that the data is fitted with the single
parameter, $C=88$ for $\alpha=400$.  The total wave action, $N=405$,
is computed from the simulations.  As for the cutoff mode, we use $k_M =
270$, which works better than the damping cutoff, $k_d = 256$.  Taking
into account that damping is a smooth function of $k$, non-zero at
$k\gtrsim k_d$, we find this modification acceptable.  Once we have
determined parameter $C$ in Eqs.\eqref{eq:Malkin1}-\eqref{eq:Malkin2},
we can combine these equations to describe the shape of the spectrum
in a parametric representation shown with solid curve in
Fig.~\ref{fig:spectra_malkin}, right panel.  To provide an explicit
expression for $n_k(k)$, one can approximate Eq.~\eqref{eq:Malkin3} as
$p_{\rm approx}(m) = \frac{3}{2}(1-m)^\frac{2}{3}$,
which gives
\begin{equation}
  \frac{n_k k^2}{k^2_{\min}} =\frac{C}{2\pi} \ln^\frac{1}{3} \left[
        1 - \left(\frac{2C}{3N} \ln\frac{k_d}{k} \right)^{\!\frac{3}{2}} \right].
\label{eq:Malkin_approx}
\end{equation}
In the range $k\in[16,256]$ where the original theory agrees with the
data, the approximation works as well as the original parametric
representation.  The agreement of the final expression with the data
might not be impressive on its own (for example, $n_kk^2 \propto
\ln^\frac{1}{2}(k_M/k)$ also provides a good fit), however the
agreement of underlying arguments gives additional support to the
theory.

Our results can be extended beyond the single value of $\alpha$, by
taking into account $n_k \propto \alpha^{\frac{1}{2}}$ scaling, which
leads to $C= c\alpha^{1/2}$ with $c \approx 4.4$.  To connect
$C\propto (n_k)^{\frac{2}{3}}$ to the energy flux, we recall our
earlier observation, $P = \nabla^2 Q \propto \alpha$, illustrated in
Fig.~\ref{fig:qflux}, to conclude that $C\propto P^{1/3}$, as proposed
by Malkin.  Yet nonlinearity restricts the range of applicability of
the theory to the high-$k$ part of the spectrum.  We speculate that the
deviation at small $k$ can be reduced, and the range of applicability
could be extended, if the comparison were done for smaller $\alpha$.

It is instructive to compare briefly the 2D case considered here with
the 3D case treated within the weak turbulence limit in
\cite{Zakharov72,Irina}. The inverse cascade is local in 3D
\cite{ZakharovLvovFalkovich1992,Zakharov72}, while the spectrum of the
direct cascade contains a logarithmic factor $\log^{-2/3}(kl_p)$
\cite{Irina}.



To conclude, we have derived and confirmed exact fourth-order flux
relations for both direct and inverse cascades. We have found the
second moments and the spectra to be nonlocal for both cascades.  The
inverse cascade is weakly turbulent at low pumping rates, while
nonlinearity is found to be substantial for the direct cascade at any
input rate and for the inverse cascade at high input rates.

We wish to stress the general lesson: an exact flux relation, which
contains neither pumping nor damping scale, must exist in the inertial
interval of every turbulence. Such relation does not preclude
turbulence from being non-local, as is the case for the direct cascade
in 2D Navier-Stokes turbulence \cite{FL94}, and as is shown here for
both cascades in 2D Gross-Pitaevsky model.




A part of this work was done during a visit to the Kavli Institute for
Theoretical Physics, UCSB, supported by grant no. NSF PHY11-25915.
N.V. was in part supported by NSF grant no. DMS-1412140.  Simulations
were performed at the Center for Advanced Research Computing (CARC),
UNM, and the Texas Advanced Computing Center (TACC) using the Extreme
Science and Engineering Discovery Environment (XSEDE), which is
supported by NSF grant no. ACI-1053575. G.F.'s work is supported by
grants of the Bi-National Science Foundation, Minerva Foundation with
funding from the German Ministry for Education and Research and by
Russian Science Foundation projects 14-22-00259 and 14-50-00150 (for
the development of the analytical theory and writing the paper).


\bibliography{cascade}


\end{document}